\documentclass[prl,twocolumn,superscriptaddress]{revtex4-1}
\usepackage[dvips]{graphicx}
\usepackage{type1cm}
\usepackage[usenames,svgnames]{xcolor}
\usepackage{color}
\usepackage{url}
\usepackage{natbib}
\usepackage[colorlinks=true,linkcolor=blue,citecolor=blue,breaklinks=true]{hyperref}

\usepackage{amsmath,amssymb,bm,amsthm}
\usepackage{newtxtext,newtxmath}

\usepackage{braket}

\DeclareMathOperator{\sgn}{\mathrm{sgn}}
\newcommand{\del}{\partial}
\newcommand{\sectionprl}[1]{{\par\it #1.---}}

\begin{document}
\nocite{apsrev41control}

\title{Heating Rates under Fast Periodic Driving beyond Linear Response}

\author{Takashi Mori}
\affiliation{
RIKEN Center for Emergent Matter Science (CEMS), Wako 351-0198, Japan
}

\begin{abstract}
Heating under periodic driving is a generic nonequilibrium phenomenon, and it is a challenging problem in nonequilibrium statistical physics to derive a quantitatively accurate heating rate. In this work, we provide a simple formula on the heating rate under fast and strong periodic driving in classical and quantum many-body systems. The key idea behind the formula is constructing a time-dependent dressed Hamiltonian by moving to a rotating frame, which is found by a truncation of the high-frequency expansion of the micromotion operator, and applying the linear-response theory. It is confirmed for specific classical and quantum models that the second-order truncation of the high-frequency expansion yields quantitatively accurate heating rates beyond the linear-response regime. Our result implies that the information on heating dynamics is encoded in the first few terms of the high-frequency expansion, although heating is often associated with an asymptotically divergent behavior of the high-frequency expansion.
\end{abstract}
\maketitle


\sectionprl{Introduction}
Fast periodic driving in thermally isolated many-body systems can stabilize interesting many-body states.
Theoretically, by using the high-frequency expansion of the Floquet operator, we can obtain a static effective Hamiltonian that describes the property of such a stabilized state, and interesting phases have been theoretically predicted~\cite{Bukov2015_review, Eckardt2017_review, Oka2019_review}.
Recent experimental developments allow us to realize some of those nonequilibrium phases~\cite{Zenesini2009, Aidelsburger2013, Miyake2013, Jotzu2014, Grushin2014, Meinert2016} and have triggered active research on the Floquet engineering (see Ref.~\cite{Oka2019_review} for a recent review).

Stability of such nonequilibrium phases is limited by heating due to periodic driving.
It is therefore practically important to evaluate the heating rate.
\emph{The rigorous approach} has made significant progress in this field.
It is rigorously proved that the heating is exponentially suppressed at high frequencies~\cite{Abanin2015, Mori2016b, Kuwahara2016, Abanin2017, Abanin2017a}.
This phenomenon is known as the Floquet prethermalization~\cite{Mori2018_review}, which has also been experimentally observed~\cite{Rubio-Abadal2020, Peng2021}.
However, the rigorous approach so far is limited to quantum systems with a bounded energy spectrum (i.e. quantum spin systems) and classical spin systems~\cite{Mori2018}.
Moreover, those rigorous results just give relatively loose upper bounds on the heating rate.

It is a theoretical challenge in nonequilibrium statistical physics to give a quantitatively accurate estimation of the heating rate for a wider class of many-body systems.
For this purpose, \emph{the statistical approach} is promising, in which the heating rate is evaluated by investigating the statistical probability of many-body resonances~\cite{Rajak2018, Rajak2019, DallaTorre2021}.
This approach is not rigorous, but instead, it gives approximate heating rates for generic systems including unbounded quantum and classical systems.
Indeed, Floquet prethermalization in classical systems has been first established along this approach~\cite{Rajak2018}.
The evaluation of the heating rate using Fermi's golden rule for quantum systems~\cite{Mallayya2019, Rakcheev2020} and the energy diffusion theory for classical systems~\cite{Hodson2021} can be categorized to this approach.
However, such treatments have successfully given accurate heating rates only for weak driving.
For modest or strong driving, which is needed for Floquet engineering, we need a new theoretical idea to achieve the goal.

In this Letter, we develop the statistical approach to the heating dynamics \emph{under strong driving}, and obtain a simple analytical formula on the heating rate.
The formula is obtained by finding a rotating frame in which driving looks weak enough. 
Such a rotating frame is found by using the technique of the high-frequency expansion ~\cite{Bukov2015_review, Eckardt2017_review, Mikami2016}.
The Hamiltonian in the rotating frame is called the \emph{dressed Hamiltonian}, which is still time-periodic but has much weaker driving amplitude.
Consequently, the linear response argument is valid for the dressed Hamiltonian rather than for the bare Hamiltonian, even when the original driving field is strong enough to being out of the linear response regime.

In a recent work~\cite{Ikeda2021}, Fermi's golden rule is extended to strong driving by utilizing the high-frequency expansion, which is conceptually close to the present work.
However, our formulation importantly differs from the one in Ref.~\cite{Ikeda2021}.
The heating-rate formula obtained in Ref.~\cite{Ikeda2021} requires the exact Floquet operator (i.e. the time evolution operator over a cycle), which is not desirable feature.
As a practical problem, this fact prevents us from applying the theory to classical systems, in which the Floquet operator is not accessible even numerically~\cite{Mori2018}.
On the other hand, the formula given in this Letter does not refer to the exact Floquet operator: the formula is completely written in terms of a truncated high-frequency expansion, which is accessible even for classical systems~\cite{Mori2018, Higashikawa2018}.
This fact tells us that information on heating under fast and strong driving is encoded in the first few terms of the high-frequency expansion, which is an important theoretical observation not found in the previous studies.

In the following, we first describe how to get a dressed Hamiltonian via the high-frequency expansion.
We then give linear response formulae on the heating rate in terms of the dressed Hamiltonian.
Next, we numerically evaluate heating rates in specific classical and quantum spin systems and compare them with our theoretical predictions.
Finally, we conclude our work with some remarks and future prospects.

\sectionprl{Dressed Hamiltonian}
For notational simplicity, we first focus on quantum systems, and later discuss classical systems.
Suppose a quantum system with a time-periodic Hamiltonian $H(t)=H_0+V(t)$ with $V(t)=\sum_{m=-\infty}^\infty V_me^{-im\omega t}$ with $V_0=0$ and $V_{-m}=V_m^\dagger$ (in classical systems $\dagger$ should be interpreted as the complex conjugate), where the frequency is denoted by $\omega$ and the period is given by $T=2\pi/\omega$.
The Floquet theorem states that the time evolution operator $U_{t,t_0}$ from time $t_0$ to $t$ is expressed as $U_{t,t_0}=e^{-iK(t)}e^{-iH_F(t-t_0)/\hbar}e^{iK(t_0)}$, where a time-periodic Hermitian operator $K(t)=K(t+T)$ is called the micromotion operator or the kick operator, and $H_F$ is called the Floquet Hamiltonian~\cite{Bukov2015_review}.
It is noted that the choice of $K(t)$ and $H_F$ is not unique: Defining $e^{iK'(t)}=U^\dagger e^{iK(t)}$ and $H_F'=U^\dagger H_F U$ for any time-independent unitary operator $U$, we find $U_{t,t_0}=e^{-iK'(t)}e^{-iH_F'(t-t_0)}e^{iK'(t_0)}$.
In the high-frequency limit $\omega\to\infty$, $K(t)$ becomes constant, and hence we require $\lim_{\omega\to\infty}K(t)=0$ for convenience.

Since $U_{t,t_0}$ satisfies $i\hbar dU_{t,t_0}/dt=H(t)U_{t,t_0}$, the micromotion operator and the Floquet operator are related with each other via the equality
\begin{equation}
H_F=e^{iK(t)}\left[H(t)-i\hbar\frac{d}{dt}\right]e^{-iK(t)}.
\end{equation}
This equation is interpreted as follows.
Let us move to the ``rotating frame'' associated with the unitary transformation $e^{iK(t)}$.
The Schr\"odinger equation $i\hbar d\ket{\psi(t)}/dt=H(t)\ket{\psi(t)}$ is transformed to
\begin{equation}
i\hbar \frac{d\ket{\psi'(t)}}{dt}=H_F\ket{\psi'(t)},
\end{equation}
where $\ket{\psi'(t)}=e^{iK(t)}\ket{\psi(t)}$ is the quantum state in the rotating frame.
That is, the Hamiltonian in the rotating frame is given by $H_F$, and the time dependence of the Hamiltonian is completely removed.

Although $H_F$ contains full information on the long-time evolution including the heating rate, $H_F$ is highly nonlocal and quite complicated in many-body systems~\cite{DAlessio2014, Lazarides2014} and it is difficult to extract dynamical properties from $H_F$.
It is also a hard task to numerically obtain $K(t)$ and $H_F$ exactly.

For fast driving, we can construct high-frequency expansions of $K(t)$ and $H_F$, which are accessible analytically and numerically. 
Because of the non-uniqueness of $K(t)$ and $H_F$, there are various high-frequency expansions~\cite{Mikami2016}.
In this work, we focus on the van Vleck expansion~\cite{Eckardt2015, Bukov2015_review} because of its analytical simplicity.
The formulation given below is also applicable to other high-frequency expansions such as the Floquet-Magnus expansion~\cite{Kuwahara2016, Mori2016b}.

The van Vleck high-frequency expansion is formally written in the following form:
\begin{equation}
K(t)=\sum_{k=1}^\infty\frac{\Lambda_k(t)}{\omega^k}, \quad H_F=H_0+\sum_{k=1}^\infty\frac{\Omega_k}{\omega^k}.
\end{equation}
The first two terms of the expansions are explicitly given by
\begin{align}
i\hbar\Lambda_1(t)=-\sum_{m\neq 0}\frac{V_m}{m}e^{-im\omega t}, & &
\Omega_1=\sum_{m\neq 0}\frac{[V_{-m},V_m]}{2m\hbar}
\end{align}
and
\begin{equation}
\left\{
\begin{aligned}
&i\hbar^2\Lambda_2(t)=\sum_{m\neq 0}\left(\frac{[V_m,H_0]}{m^2}+\sum_{n\neq 0,m}\frac{[V_n,V_{m-n}]}{mn}\right)e^{-im\omega t},\\
&\Omega_2=\sum_{m\neq 0}\left(\frac{[[V_{-m},H_0],V_m]}{2\hbar^2m^2}+\sum_{n\neq 0,m}\frac{[[V_{-m},V_{m-n}],V_n]}{3\hbar^2mn}\right).
\end{aligned}
\right.
\end{equation}
Additional details on the van Vleck expansion as well as an explicit form of $\Lambda_3(t)$ are given in Supplementary Material (SM)~\cite{SM} (also see Ref.~\cite{Mikami2016}).

It should be noted that this expansion is an asymptotic expansion in the thermodynamic limit~\cite{Mori2016b}, and hence we should truncate the expansion to obtain a meaningful result.
We define the $n$th order truncation of the expansion of $K(t)$ and $H_F$ as $K^{(n)}(t)=\sum_{k=1}^n\Lambda_k/\omega^k$ and $H_F^{(n)}=\sum_{k=0}^n\Omega_k/\omega^k$, respectively.

Let us move to the rotating frame associated with $K^{(n)}(t)$ rather than $K(t)$.
The Hamiltonian in this rotating frame is given by
\begin{align}
e^{-iK^{(n)}(t)}\left[H(t)-i\hbar\frac{d}{dt}\right]e^{iK^{(n)}t}\simeq\tilde{H}^{(n)}(t)
\label{eq:H_rotating}
\end{align}
up to $O(\omega^{-n})$, where the $n$th order \emph{dressed Hamiltonian} $\tilde{H}^{(n)}$ is given by
\begin{equation}
\tilde{H}^{(n)}(t)=H_F^{(n)}+V^{(n)}(t).
\end{equation}
Its static part is nothing but the $n$th order truncation of the Floquet Hamiltonian.
In SM~\cite{SM}, it is shown that the dressed driving $V^{(n)}(t)$ is expressed as
\begin{equation}
V^{(n)}(t)=\frac{\hbar}{\omega^{n+1}}\frac{d\Lambda_{n+1}(t)}{dt}.
\label{eq:Vn}
\end{equation}
We find that $V^{(n)}(t)$ satisfies the desirable property of periodic driving: $V^{(n)}(t)=V^{(n)}(t+T)$ and $\int_0^Tdt\, V^{(n)}(t)=0$.
Moreover, the dressed driving field is strongly weakened at high frequencies: the amplitude of $V^{(n)}(t)$ is smaller by a factor of $(g/\hbar\omega)^n$ compared with the bare driving field $V(t)$, where $g$ denotes a characteristic local energy scale of the Hamiltonian $H(t)$~\footnote{For strong driving with amplitude $\xi$ much larger than any other local energy scale, we have $g\sim\xi$.}.
It is therefore expected that even if the driving is strong in the original frame, it looks weak in the rotating frame, and we can carry out the linear response calculation in the latter.

We remark that higher-order terms omitted in Eq.~(\ref{eq:H_rotating}) are smaller by a factor of $(g/\hbar\omega)^{n+1}$, and hence Eq.~(\ref{eq:H_rotating}) is justified when $(g/\hbar\omega)^{n+1}\ll 1$.
On the other hand, our approximation breaks down when $g\gtrsim\hbar\omega$ (e.g. when the amplitude $\xi$ of periodic driving is greater than $\hbar\omega$).
 
The dressed Hamiltonian can also be constructed in classical systems.
Analytical expressions of $\Lambda_k(t)$ and $\Omega_k$ in quantum systems contain commutators of operators, and their classical counterparts are just obtained by replacing the commutator by the Poisson bracket, $(1/i\hbar)[\cdot,\cdot]\to \{\cdot,\cdot\}_\mathrm{PB}$.
This procedure is justified by formally applying Floquet theory to the classical Liouville equation $d\rho(z_t)/dt=\{H(t;z_t),\rho(z_t)\}_\mathrm{PB}$, where $z_t$ represents the set of all the coordinates and all the momenta of the classical system at time $t$, $\rho(z_t)$ is the probability distribution in the classical phase space, and $H(t;z_t)$ denotes the classical Hamiltonian.
See Ref.~\cite{Mori2018} for more details.

\sectionprl{Heating-rate formula}
We now give the heating rate by using the linear response theory for the dressed Hamiltonian $\tilde{H}^{(n)}(t)=H_F^{(n)}+V^{(n)}(t)$.
Now we interpret the static part $H_F^{(n)}$ of $\tilde{H}^{(n)}(t)$ as the energy of the system, and the heating rate is given by $\kappa=d(E/N)/dt$, where $N$ denotes the system size (the number of particles/spins).
Here, $V^{(n)}(t)$ is decomposed as $V^{(n)}(t)=\sum_{m=-\infty}^\infty V_m^{(n)}e^{-im\omega t}$.

According to the linear response theory~\cite{Kubo_text, Kubo1957}, the heating rate $\kappa$ at the energy $E$ under the external field $V^{(n)}(t)$ is evaluated in terms of auto-correlation functions of $\{\dot{V}^{(n)}_m\}$, where for classical systems $\dot{V}^{(n)}_m=\{V^{(n)}_m,H_F^{(n)}\}_\mathrm{PB}$ and for quantum systems $\dot{V}^{(n)}_m=(1/i\hbar)[V^{(n)}_m,H_F^{(n)}]$.
The formula is given by
\begin{equation}
\kappa=\frac{\beta}{2N}\sum_{m\neq 0}C_m(\omega),
\label{eq:heating}
\end{equation}
where $\beta=\del S(E)/\del E$ denotes the microcanonical temperature [$S(E)$ is the microcanonical entropy]. 
For classical systems, the function $C_m(\omega)$ is defined as
\begin{equation}
C_m(\omega)=\int_{-\infty}^\infty dt\;\braket{\dot{V}^{(n)\dagger}_m(z_t)\dot{V}^{(n)}_m(z_0)}e^{im\omega t},
\label{eq:C_classical}
\end{equation}
where $\braket{\cdot}$ denotes the microcanonical average and the trajectory $\{z_t\}_t$ is generated by the static Hamiltonian $H_F^{(n)}$, i.e., $dz_t/dt=\{z_t,H_F^{(n)}(z_t)\}_\mathrm{PB}$.
For quantum systems,
\begin{align}
C_m(\omega)&=\int_{-\infty}^\infty dt\frac{1}{\beta}\int_0^\beta d\lambda\; \braket{\dot{V}^{(n)\dagger}_m(t-i\hbar\lambda)\dot{V}^{(n)}_m(0)}e^{im\omega t}
\nonumber \\
&=\frac{1-e^{-\beta m\omega}}{\beta m\omega}\int_{-\infty}^\infty dt\braket{\dot{V}^{(n)\dagger}_m(t)\dot{V}^{(n)}_m(0)}e^{im\omega t},
\label{eq:C_quantum}
\end{align}
where $\dot{V}^{(n)\dagger}_m(t)=e^{(i/\hbar)H_F^{(n)}t}\dot{V}^{(n)}_me^{-(i/\hbar)H_F^{(n)}t}$.
The derivation of Eqs.~(\ref{eq:heating}), (\ref{eq:C_classical}) and (\ref{eq:C_quantum}) is given in SM~\cite{SM}.

The Wiener-Khinchin theorem states that $C_m(\omega)$ is identical to the power spectrum of $\dot{V}_m^{(n)}(z_t)$ for classical systems~\cite{Kubo_text}.
It is also extended to quantum systems; see Ref.~\cite{Ikeuchi2015} for the calculation of the Fourier transform of auto-correlation functions via a Wiener-Khinchin-like theorem for quantum systems.

When the frequency is large enough, $C_m(\omega)$ decays exponentially in $|m\omega|$~\cite{Abanin2015, Hodson2021}.
Therefore, the contribution from $m=\pm 1$ is dominant, and the heating rate is approximated by
\begin{equation}
\kappa\simeq\frac{\beta}{2}[C_1(\omega)+C_{-1}(\omega)].
\label{eq:heating_approx}
\end{equation}
Equations~(\ref{eq:heating}) and (\ref{eq:heating_approx}) are our main result.

It should be noted that when $n=0$, our formula is reduced to the conventional linear response result.
For $n\geq 1$, our result is regarded as its extension to fast and strong driving.
It is expected that increasing $n$ improves accuracy up to a certain order $n_0\propto\hbar\omega/g$, but increasing $n$ further for $n>n_0$ will be rather harmful because of the divergence of the high-frequency expansion~\cite{Kuwahara2016, Mori2016b}.

Before going on to numerical results, it should be emphasized that it is crucial in our formulation to consider the high-frequency expansion of the \emph{micromotion operator}.
Since the micromotion operator describes fast oscillations rather than long-time slow dynamics, it is often neglected.
However, in our formulation, a truncation of the high-frequency expansion of the micromotion operator yields a dressed driving field $V^{(n)}(t)$, which contributes to a finite heating rate. 

\sectionprl{Classical model}
We now present numerical results for classical spin systems.
The classical spin at $i$th site is denoted by $\bm{s}_i=(s_i^x,s_i^y,s_i^z)$ satisfying $|\bm{s}_i|^2=1$, and $z=(\bm{s}_1,\bm{s}_2,\dots,\bm{s}_N)$ represents the set of all the classical spin variables.
The Hamiltonian $H(t;z)=H_0(z)+V(t;z)$ is given by
\begin{equation}
H_0(z)=-\sum_{i=1}^N\left[Js_i^zs_{i+1}^z+h_xs_i^x+h_zs_i^z\right]
\end{equation}
and
\begin{equation}
V(t;z)=-\xi\left[\cos(\omega t)\sum_{i=1}^Ns_i^zs_{i+1}^z+\sin(\omega t)\sum_{i=1}^Ns_i^x\right].
\end{equation}
By defining the local effective field $\tilde{\bm{h}}_i(t)=-\del H(t)/\del\bm{s}_i$, the classical equations of motion are given by
\begin{equation}
\frac{d\bm{s}_i}{dt}=2\bm{s}_i\times\tilde{\bm{h}}_i(t),
\label{eq:EOM}
\end{equation}
which is the classical limit of the Heisenberg equations of motion for Pauli matrices.
In this work, we fix $J=1$, $h_x=0.77$, $h_z=0.49$, $T=0.5$ ($\omega=2\pi/T\simeq 12.6$), and the system size $N=100$. 

The heating rate is calculated as follows.
First $s_i^y$ and $s_i^z$ for $i=1,2,\dots,N$ are sampled independently from the uniform distribution between 0 and 0.1, and $s_i^x$ is fixed as $s_i^x=\sqrt{1-s_i^y(0)^2-s_i^z(0)^2}$.
We then randomly choose $\tau\in[1000,2000]$ and the spin variables evolve over the time $\tau$ without driving, $\xi=0$.
This is our initial state $\{\bm{s}_i(0)\}_{i=1}^N$.
Next, we solve Eq.~(\ref{eq:EOM}).
We measure two times $t_1$ and $t_2$, which are defined as $t_i=\min_{n\in\mathbb{N}}\{t=nT: H_0(z_t)\geq N\varepsilon_i\}$ for $i=1,2$.
We fix $\varepsilon_1=-0.6$ and $\varepsilon_2=-0.5$ (the corresponding inverse temperature is $\beta\simeq 1.1$).
The heating rate is then given by $\kappa=(\varepsilon_2-\varepsilon_1)/(t_2-t_1)$.
We repeat this procedure 500 times, and compute the average heating rate.

We compare it with the heating rate calculated by our formula.
We perform the van Vleck high-frequency expansion and analytically obtain $\tilde{H}^{(n)}(t)$ up to $n=2$.
We then evaluate the heating rate by using Eq.~(\ref{eq:heating_approx}) with the help of the Wiener-Khinchin theorem.
The technical detail of the calculation is explained in SM~\cite{SM}.

\begin{figure}[t]
\centering
\includegraphics[width=1\linewidth]{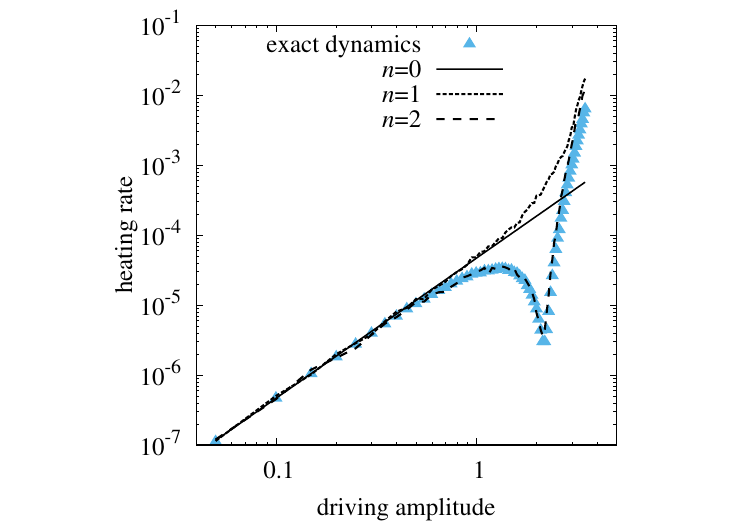}
\caption{Heating rate in the classical spin system against the driving amplitude $\xi$.
The system size is set to be $N=100$.
Blue triangles show the heating rate estimated by exactly solving the equations of motion~(\ref{eq:EOM}).
The solid line, the dotted line, and the dashed line show the heating rates calculated by our formula~(\ref{eq:heating_approx}) for $n=0$, 1, and 2, respectively.
Error bars are smaller than the symbols.}
\label{fig:heating_c}
\end{figure}

Our numerical results are displayed in Fig.~\ref{fig:heating_c}.
The heating rate calculated by solving Eq.~(\ref{eq:EOM}) shows non-monotonic behavior: the heating is suppressed for strong driving~\cite{Ikeda2021}.
On the other hand, for $n=0$ and 1, our formula~(\ref{eq:heating_approx}) does not reproduce non-monotonicity.
When $n=0$, our formula is reduced to the linear response expression, and hence the heating rate is proportional to $\xi^2$.
When $n=1$, our formula agrees with the exact heating rate at weak and strong driving, but does not show non-monotonicity.
We clearly see that our formula for $n=2$ well reproduces a curve of the exact heating rate, including characteristic non-monotonic behavior.

Frequency dependences of the heating rate are given in SM~\cite{SM}, in which we find that the non-monotonicity occurs at $\xi$ which is large but independent of $\omega$.
Therefore, this non-monotonicity should be distinguished from dynamical freezing phenomena~\cite{Das2010, Haldar2018, Haldar2021}, in which heating is suppressed due to an emergent symmetry at ultra-strong driving with $\xi\propto\omega$.

\begin{figure}[t]
\centering
\includegraphics[width=1\linewidth]{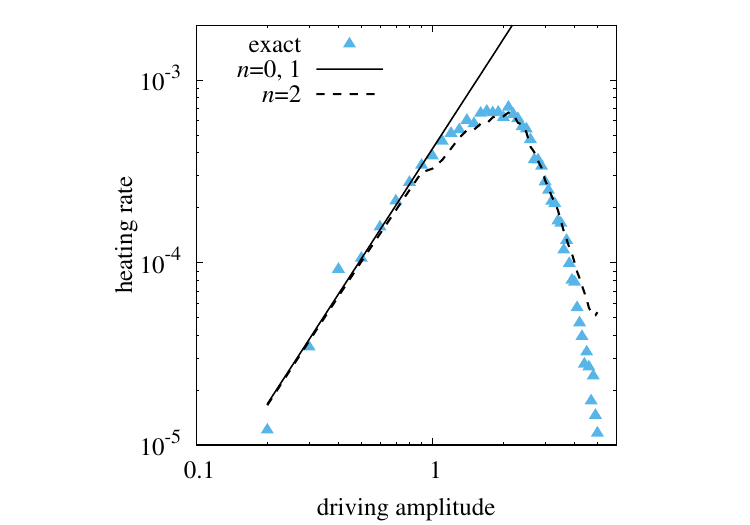}
\caption{Heating rate in the quantum spin system against the driving amplitude $\xi$.
Blue triangles show the heating rate estimated by exactly solving the Scr\"odinger equation for $N=16$.
The solid line and the dashed line show the heating rates calculated by our formula~(\ref{eq:heating_approx}) for $N=14$ with $n=0$ and 2, respectively (in the present model, $n=1$ gives an identical result to $n=0$).
Error bars are smaller than the symbols.}
\label{fig:heating_q}
\end{figure}

\sectionprl{Quantum model}
We also verify our formula in quantum systems.
We consider a quantum spin-1/2 chain with the Hamiltonian
\begin{equation}
\left\{
\begin{aligned}
&H_0=-\sum_{i=1}^N[J_z\sigma_i^z\sigma_{i+1}^z+J_x\sigma_i^x\sigma_{i+1}^x+h\sigma_i^z], \\
&V(t)=-\xi\sgn[\cos(\omega t)]\sum_{i=1}^N\sigma_i^x,
\end{aligned}
\right.
\end{equation}
where $\sigma^\alpha$ ($\alpha=x,y,z$) denotes the Pauli matrix.
We fix $J_z=1$, $J_x=0.77$, $h=0.6$, and $T=0.5$ ($\omega\simeq 12.6$).

We prepare an initial state as a canonical thermal pure quantum state~\cite{Sugiura2013}: we generate a random vector $\ket{r}$ whose elements are i.i.d. Gaussian of mean 0 and unit variance, and construct a state $\ket{\beta}=e^{-\beta H/2}\ket{r}/\braket{r|e^{-\beta H}|r}$.
We then solve the Schr\"odinger equation $i\hbar d\ket{\psi(t)}/dt=H(t)\ket{\psi(t)}$ with $\ket{\psi(0)}=\ket{\beta}$, where we set $\hbar=1$ in numerical calculations.
The heating rate is calculated in the same way as in classical systems: we measure $t_1$ and $t_2$ satisfying $t_i=\min_{n\in\mathbb{N}}\{t=nT: \braket{\psi(t)|H_0|\psi(t)}\geq N\varepsilon_i\}$ with $\varepsilon_1=-0.5$ and $\varepsilon_2=-0.48$ (the corresponding inverse temperature is $\beta\simeq 0.23$).
The heating rate is given by $(\varepsilon_2-\varepsilon_1)/(t_2-t_1)$.
For the system size $N=16$, we repeat the above procedure (the generation of an initial state, solving the Schr\"odinger equation, and measuring the heating rate) 10 times, and compute the average heating rate.

The heating rate is also evaluated for $N=14$ by using our formula with $n=0$ and 2 (in the present model, it is shown that $n=1$ gives the identical result to $n=0$).
Details are explained in SM~\cite{SM}.

Numerical results are shown in Fig.~\ref{fig:heating_q}.
We can see that the heating rate again shows non-monotonic behavior, which implies that the system is not in the linear response regime. 
Our formula with $n=2$ reproduces this behavior.

\sectionprl{Conclusion and Outlook}
We have derived a formula on the heating rate under fast driving with arbitrary driving strength.
Our idea is based on considering the problem in a rotating frame in which driving looks weak.
Such a rotating frame is found by using the high-frequency expansion of the micromotion operator.
Our formulation is valid for both classical and quantum systems.

It is often argued that a truncation of the high-frequency expansion of the Floquet Hamiltonian describes dynamics in a prethermal regime before the heating takes place~\cite{Mori2016b, Kuwahara2016, Ikeda2021}, whereas an asymptotic divergent behavior of the high-frequency expansion is related to heating.
Contrary to this argument, our formulation tells us that the information on heating under fast and strong periodic driving is encoded in a truncation of the high-frequency expansions of the Floquet Hamiltonian \emph{and} the micromotion operator.
Considering the micromotion operator is crucial in our formulation, although it is often neglected in investigating heating dynamics because of the fact that the micromotion describes fast oscillations rather than long-time slow dynamics.

Both in classical and quantum systems, we have found non-monotonic heating rates as a function of the driving amplitude.
Such non-monotonicity has also been found in the previous study~\cite{Ikeda2021}, and it looks universal in some extent.
It is a future problem to understand universal features of the heating dynamics by using our formulation.

Some recent studies have also attempted to use aperiodic driving (random or quasiperiodic one) for controlling quantum many-body systems~\cite{Nandy2017, Zhao2019, Crowley2020, Zhao2021, Long2021}, and some rigorous results have begun to appear~\cite{Else2020, Mori2021}.
It will be a fascinating open problem to give a simple and accurate heating-rate formula for fast and strong quasiperiodic driving.

\begin{acknowledgments}
Fruitful discussions with Wade Hodson, Tatsuhiko N. Ikeda, and Christopher Jarzynski are gratefully acknowledged.
This work was supported by JSPS KAKENHI Grant Numbers JP19K14622, JP21H05185.
\end{acknowledgments}

\bibliography{apsrevcontrol,physics,SM}

\end{document}


\nocite{apsrev41control}

\title{Supplemental Material}

\author{Takashi Mori}
\affiliation{
RIKEN Center for Emergent Matter Science (CEMS), Wako 351-0198, Japan
}

\setcounter{equation}{0}
\def\theequation{S\arabic{equation}}
\setcounter{figure}{0}
\def\thefigure{S\arabic{figure}}
\setcounter{section}{0}
\def\thesection{\Alph{section}}
\renewcommand*{\citenumfont}[1]{S#1}
\renewcommand*{\bibnumfmt}[1]{[S#1]}

\maketitle

\section{Derivation of the linear response formula on the heating rate}

In this section, we give linear response formulae on the heating rate.
The material in this section is rather standard and not new, but we provide it for convenience.

\subsection*{Quantum systems}
Let us consider the Hamiltonian $H(t)=H_0+V(t)$ with $V(t)=\sum_{m\neq 0}V_me^{-im\omega t}$.
We consider a quantum system below, but it is straightforward to extend the analysis to a classical system.

Let us define $\Phi_{mn}(t)$ as the response function of the quantity $V_m$ to the external field proportional to $V_n$.
That is, when $V(t)$ is small, the expectation value of $V_m$ at time $t$ is written as
\begin{equation}
\braket{V_m}_t-\braket{V_m}_\mathrm{eq}=\sum_{n\neq 0}\int_0^\infty ds\, \Phi_{mn}(s)e^{-in\omega(t-s)}
=\sum_{n\neq 0}e^{-in\omega t}\int_0^\infty ds\, \Phi_{mn}(s)e^{in\omega s}
=:\sum_{n\neq 0}e^{-in\omega t}\chi_{mn}(n\omega),
\end{equation}
where $\braket{\cdot}_\mathrm{eq}$ denotes the equilibrium average without perturbation $V(t)$.

The energy absorption rate $\kappa$ is given by
\begin{equation}
\kappa=\frac{1}{N}\sum_{m\neq 0}\overline{e^{-im\omega t}\frac{d\braket{V_m}_t}{dt}}=\frac{1}{N}\sum_{m\neq 0}im\omega\chi_{m,-m}(-m\omega)=-\frac{i}{N}\sum_{m\neq 0}m\omega\chi_{-m,m}(m\omega),
\end{equation}
where $\overline{f(t)}=(1/T)\int_0^Tdt\, f(t)$ denotes the time average.

According to the linear response theory~\cite{Kubo1957}, $\Phi_{nm}(t)$ is given by
\begin{equation}
\Phi_{mn}(t)=\int_0^\beta d\lambda\, \braket{\dot{V}_n(-i\hbar\lambda)V_m(t)}.
\end{equation}
Therefore, we have
\begin{equation}
\kappa=-\frac{1}{N}\sum_{m\neq 0}im\omega\int_0^\infty dt\int_0^\beta d\lambda\, \braket{\dot{V}_m(-i\hbar\lambda)V_m^\dagger(t)}e^{im\omega t}.
\end{equation}
By integrating by part, we obtain
\begin{equation}
\kappa=\frac{1}{N}\sum_{m\neq 0}\int_0^\infty dt\int_0^\beta d\lambda\, \braket{\dot{V}_m(-i\hbar\lambda)\dot{V}_m^\dagger (t)}e^{im\omega t}.
\end{equation}
By taking the complex conjugate and changing the integration variable $t\to -t$, we have
\begin{equation}
\int_0^\infty dt\int_0^\beta d\lambda\, \braket{\dot{V}_m(-i\hbar\lambda)\dot{V}_m^\dagger (t)}e^{im\omega t}=\int_{-\infty}^0dt\int_0^\beta d\lambda\,\braket{\dot{V}_m(-t)\dot{V}_m^\dagger(i\hbar\lambda)}e^{im\omega t}
=\int_{-\infty}^0dt\int_0^\beta d\lambda\,\braket{\dot{V}_m(-i\hbar\lambda)\dot{V}_m^\dagger(t)}e^{im\omega t}.
\end{equation}
We thus obtain
\begin{equation}
\kappa=\frac{1}{2N}\sum_{n\neq 0}\int_{-\infty}^\infty dt\int_0^\beta d\lambda\, \braket{\dot{V}_m(-i\hbar\lambda)\dot{V}_m^\dagger (t)}e^{im\omega t}.
\end{equation}
By changing $t\to -t$ and $m\to -m$, $\kappa$ is also written as
\begin{equation}
\kappa=\frac{1}{2N}\sum_{m\neq 0}\int_{-\infty}^\infty dt\int_0^\beta d\lambda\, \braket{\dot{V}_m^\dagger(t-i\hbar\lambda)\dot{V}_m(0)}e^{im\omega t}.
\label{eq:quantum_formula}
\end{equation}
This is identical to the formula given in the main text.

Let us write $H_0=\sum_a E_a\ket{a}\bra{a}$, and the equilibrium density matrix $\rho_\mathrm{eq}=\sum_a p_a\ket{a}\bra{a}$ with $p_a\geq 0$ and $\sum_ap_a=1$.
$\kappa$ is then written as
\begin{align}
\kappa&=\frac{1}{2N}\sum_{m\neq 0}\int_{-\infty}^\infty dt\int_0^\beta d\lambda\,\sum_{a,b}e^{\lambda(E_a-E_b)}e^{(i/\hbar)(E_a-E_b+m\hbar\omega)t}\lvert\braket{b|\dot{V}_m|a}\rvert^2p_a
\nonumber \\
&=\frac{1}{2N}\sum_{m\neq 0}\sum_{a,b}\frac{1-e^{-\beta(E_b-E_a)}}{E_b-E_a}2\pi\hbar\delta(E_b-E_a-m\hbar\omega)\lvert\braket{b|\dot{V}_m|a}\rvert^2p_a
\nonumber \\
&=\frac{1}{2N}\sum_{m\neq 0}\sum_{a,b}\frac{1-e^{-\beta m\hbar\omega}}{m\hbar\omega}2\pi\hbar\delta(E_b-E_a-n\hbar\omega)\lvert\braket{b|\dot{V}_m|a}\rvert^2p_a
\nonumber \\
&=\frac{1}{2N}\sum_{m\neq 0}\frac{1-e^{-\beta m\hbar\omega}}{m\hbar\omega}\int_{-\infty}^\infty dt\, \braket{\dot{V}_m^\dagger(t)\dot{V}_m(0)}e^{im\omega t}.
\end{align}
By using $\braket{b|\dot{V}_m|a}=\frac{E_b-E_a}{i\hbar}\braket{b|V_m|a}$, we can also express $\kappa$ as
\begin{equation}
\kappa=\frac{1}{N}\frac{\pi}{\hbar}\sum_{m\neq 0}\sum_{a,b}(1-e^{-\beta m\hbar\omega})m\hbar\omega\delta(E_b-E_a-m\hbar\omega)\lvert\braket{b|V_m|a}\rvert^2p_a.
\label{eq:quantum_kappa}
\end{equation}
When $p_a$ is the canonical distribution $p_a=e^{-\beta E_a}/Z$, we have
\begin{equation}
\kappa=\frac{1}{N}\frac{\pi}{\hbar}\sum_{m\neq 0}\sum_{a,b}(p_a-p_b)m\hbar\omega\delta(E_b-E_a-m\hbar\omega)\lvert\braket{b|V_m|a}\rvert^2,
\end{equation}
which is identical to the heating rate for the canonical ensemble calculated by Fermi's golden rule.

\subsection*{Classical systems}
In a classical system, by taking the limit of $\hbar\to 0$, we obtain
\begin{equation}
\kappa=\frac{\beta}{2N}\sum_{m\neq 0}\int_{-\infty}^\infty dt\,\braket{\dot{V}_m^*(z_t)\dot{V}_m(z_0)}e^{im\omega t},
\label{eq:classical_formula}
\end{equation}
where $z_t$ is the set of all the canonical variables describing the classical system at time $t$.

By using Wiener-Khinchin theorem~\cite{Kubo_text}, the Fourier transform of the auto-correlation function
\begin{equation}
C_m(\omega)=\int_{-\infty}^\infty dt\,\braket{\dot{V}_m^\dagger(z_t)\dot{V}_m(z_0)}e^{im\omega t}
\end{equation}
is evaluated by computing the power spectrum of $\dot{V}_n(z_t)$.
Let us define
\begin{equation}
a_k=\frac{1}{\tau}\int_0^\tau dt\, A(z_t)e^{-i\omega_k t}, \quad \omega_k=\frac{2\pi k}{\tau}
\end{equation}
with $k$ an integer.
The Wiener-Khinchin theorem states that
\begin{equation}
\int_{-\infty}^\infty dt\, \braket{A^*(z_t)A(z_0)}e^{i\omega t}=\lim_{\tau\to\infty,\omega_k\to\omega}\tau \braket{|a_k|^2}.
\end{equation}
By using this relation, we can compute $C_m(\omega)$ through the power spectrum of $\dot{V}_m(z_t)$.

\section{Van Vleck high-frequency expansion}

In this section, we review the van Vleck high-frequency expansion~\cite{Bukov2015_review, Eckardt2015}.
As is explained in the main text, the Floquet Hamiltonian and the micromotion operator satisfy the equality
\begin{equation}
H_F=e^{iK(t)}\left[H(t)-i\hbar\frac{d}{dt}\right]e^{-iK(t)}.
\label{eq:equality0}
\end{equation}
Here, we rewrite $e^{iK(t)}H(t)e^{-iK(t)}$ and $e^{iK(t)}i\hbar(d/dt)e^{-iK(t)}$ as follows:
\begin{equation}
e^{iK(t)}H(t)e^{-iK(t)}=\sum_{n=0}^\infty\frac{1}{n!}\mathrm{ad}_{iK(t)}^nH(t),
\end{equation}
where $\mathrm{ad}_{A}:=[A,\cdot]$, and
\begin{align}
e^{iK(t)}i\hbar(d/dt)e^{-iK(t)}
&=e^{iK(t)}i\hbar\int_0^1dx\, e^{-iK(t)(1-x)}\left(-i\frac{dK(t)}{dt}\right)e^{-iK(t)x}
\nonumber \\
&=\hbar\int_0^1dx\,e^{iK(t)x}\frac{dK(t)}{dt}e^{-iK(t)x}
\nonumber \\
&=\hbar\int_0^1dx\,\sum_{n=0}^\infty\frac{x^n}{n!}\mathrm{ad}_{iK(t)}^n\frac{dK(t)}{dt}
\nonumber \\
&=\hbar\sum_{n=0}^\infty\frac{1}{(n+1)!}\mathrm{ad}_{iK(t)}^n\frac{dK(t)}{dt}.
\end{align}
Equation~(\ref{eq:equality0}) is thus rewritten as
\begin{equation}
H_F=\sum_{n=0}^\infty \mathrm{ad}_{iK(t)}^n\left[\frac{1}{n!}H(t)-\frac{\hbar}{(n+1)!}\frac{dK(t)}{dt}\right].
\label{eq:equality}
\end{equation}

The van Vleck high-frequency expansion corresponds to the expansion
\begin{equation}
K(t)=\sum_{k=1}^\infty \frac{\Lambda_k(t)}{\omega^k}, \quad H_F=H_0+\sum_{k=1}^\infty\frac{\Omega_k}{\omega^k}
\label{eq:expansion}
\end{equation}
with the boundary condition
\begin{equation}
\int_0^T dt\, K(t)=0.
\label{eq:boundary}
\end{equation}
By substituting Eq.~(\ref{eq:expansion}) into Eq.~(\ref{eq:equality}) and compare the left-hand side and the right-hand side order by order, we obtain an explicit expression of $\{\Lambda_k(t)\}$ and $\{\Omega_k\}$.
Here, $\Lambda_k(t)$ is periodic in time, $\Lambda_k(t)=\Lambda_k(t+T)$, and hence the time derivative in Eq.~(\ref{eq:equality}) should be treated as the order $\omega^{-1}$.

At the lowest order, we have
\begin{equation}
H_0=H_0+V(t)-i\frac{\hbar}{\omega}\frac{d}{dt}\left[-i\Lambda_1(t)\right],
\end{equation}
which yields
\begin{equation}
\frac{\hbar}{\omega}\frac{d}{dt}\Lambda_1(t)=V(t).
\label{eq:driving_1}
\end{equation}
From Eq.~(\ref{eq:boundary}) and the periodicity $\Lambda_1(t)=\Lambda_1(t+T)$ with $T=2\pi/\omega$, $\Lambda_1(t)$ is expressed as
\begin{equation}
\Lambda_1(t)=\sum_{m\neq 0}\Lambda_{1,m}e^{-im\omega t}.
\end{equation}
By substituting it into Eq.~(\ref{eq:driving_1}), we have $\Lambda_{1,m}=iV_m/(m\hbar)$ and thus
\begin{equation}
i\hbar\Lambda_1(t)=-\sum_{m\neq 0}\frac{V_m}{m}e^{-im\omega t}.
\end{equation}

At the next order $O(\omega)$, we have
\begin{equation}
\Omega_1=[i\Lambda_1(t),H_0+V(t)]-\frac{\hbar}{\omega}\frac{d\Lambda_2(t)}{dt}-\frac{\hbar}{2\omega}\left[i\Lambda_1(t),\frac{d\Lambda_1(t)}{dt}\right]
=\left[i\Lambda_1(t),H_0+\frac{1}{2}V(t)\right]-\frac{\hbar}{\omega}\frac{d\Lambda_2(t)}{dt},
\label{eq:equality_1}
\end{equation}
where Eq.~(\ref{eq:driving_1}) is used in the last equality.
Now we take the time average of the both sides of Eq.~(\ref{eq:equality_1}), which yields
\begin{equation}
\Omega_1=\frac{1}{T}\int_0^T dt\, \left[i\Lambda_1(t),\frac{V(t)}{2}\right]=\sum_{m\neq 0}\frac{[V_{-m},V_m]}{2m\hbar}.
\end{equation}
We can also determine $d\Lambda_2(t)/dt$ as
\begin{align}
\frac{\hbar}{\omega}\frac{d\Lambda_2(t)}{dt}&=[i\Lambda_1(t),H_0]+\frac{1}{2}[i\Lambda_1(t),V(t)]-\Omega_1
\nonumber \\
&=-\sum_{m\neq 0}\frac{[V_m,H_0]}{m\hbar}e^{-im\omega t}-\frac{1}{2}\sum_{m\neq 0}\sum_{n\neq 0,m}\frac{[V_n,V_{m-n}]}{n\hbar}e^{-im\omega t}.
\end{align}
By using the boundary condition $\int_0^Tdt\, \Lambda_2(t)=0$, we have
\begin{equation}
i\hbar^2\Lambda_2(t)=\sum_{m\neq 0}\frac{[V_m,H_0]}{m^2}e^{-im\omega t}+\sum_{m\neq 0}\sum_{n\neq 0,m}\frac{[V_n,V_{m-n}]}{mn}e^{-im\omega t}.
\end{equation}

By repeating the above procedure to higher order terms, we can recursively determine $\{\Omega_k\}$ and $\{\Lambda_k\}$.
For $\Omega_2$ and $\Lambda_3(t)$, we only give the final result~\cite{Mikami2016}:
\begin{equation}
\hbar^2\Omega_2=\sum_{m\neq 0}\frac{[[V_{-m},H_0],V_m]}{2m^2}+\sum_{m\neq 0}\sum_{n\neq 0,m}\frac{[[V_{-m},V_{m-n}],V_n]}{3mn},
\end{equation}
and
\begin{align}
i\hbar^3\Lambda_3(t)=&-\sum_{m\neq 0}\frac{[[V_m,H_0],H_0]}{m^3}e^{-im\omega t}+\sum_{m\neq 0}\sum_{n\neq 0}\frac{[V_m,[V_{-n},V_n]]}{4m^2n}e^{-im\omega t}
\nonumber \\
&-\sum_{m\neq 0}\sum_{n\neq 0,m}\frac{[[V_n,H_0],V_{m-n}]}{2mn^2}e^{-im\omega t}-\sum_{m\neq 0}\sum_{n\neq 0,m}\frac{[[V_n,V_{m-n}],H_0]}{2m^2n}e^{-im\omega t}
\nonumber \\
&-\sum_{m\neq 0}\sum_{n\neq 0}\sum_{l\neq 0,n,m}\frac{[[V_n,V_{l-n}],V_{m-l}]}{4mnl}e^{-im\omega t}-\sum_{m\neq 0}\sum_{n\neq 0}\sum_{l\neq 0,m-n}\frac{[V_n,[V_l,V_{m-n-l}]]}{12mnl}e^{-im\omega t}.
\end{align}

\section{Dressed Hamiltonian}

In the main text, we define the dressed Hamiltonian
\begin{equation}
\tilde{H}^{(n)}(t)=e^{iK^{(n)}(t)}\left[H(t)-i\hbar\frac{d}{dt}\right]e^{-iK^{(n)}(t)}=H^{(n)}_0+V^{(n)}(t)+O(\omega^{-(n+1)}),
\end{equation}
where $K^{(n)}(t)=\sum_{k=1}^n\Lambda_k/\omega^k$ is the $k$th order truncation of the van Vleck expansion of $K(t)$, and $V^{(n)}(t)$ satisfies $V^{(n)}(t)=V^{(n)}(t+T)$ and $\int_0^Tdt\, V^{(n)}(t)=0$.

We now prove
\begin{equation}
H_0^{(n)}=H_F^{(n)}:=\sum_{k=0}^n\frac{\Omega_k}{\omega^k},
\label{eq:H_0^n}
\end{equation}
i.e. the static part of the dressed Hamiltonian is nothing but the $n$th order truncation of the high-frequency expansion of $H_F$, and
\begin{equation}
V^{(n)}(t)=\frac{\hbar}{\omega^{n+1}}\frac{d\Lambda_{n+1}(t)}{dt}.
\label{eq:V_n}
\end{equation}

For the proof, we use the following expression corresponding to Eq.~(\ref{eq:equality}):
\begin{equation}
\tilde{H}^{(n)}(t)=\sum_{m=0}^\infty\mathrm{ad}_{iK^{(n)}(t)}^m\left[\frac{1}{m!}H(t)-\frac{\hbar}{(m+1)!}\frac{dK^{(n)}(t)}{dt}\right].
\end{equation}
By comparing it with Eq.~(\ref{eq:equality}), $\tilde{H}^{(n)}(t)$ and $H_F$ are identical up to $O(\omega^{-n})$ except for the term $-(\hbar/\omega^{n+1})d\Lambda_{n+1}(t)/dt$ (remember that the time derivative should be regarded as $O(\omega)$).
Therefore, we have
\begin{equation}
\tilde{H}^{(n)}(t)=H_F^{(n)}+\frac{\hbar}{\omega^{n+1}}\frac{d\Lambda_{n+1}(t)}{dt},
\end{equation}
which implies Eq.~(\ref{eq:V_n}).

\section{Explicit expression of the dressed Hamiltonian}
We give an explicit expression of the dressed Hamiltonian for the models discussed in the main text.

For the classical model
\begin{equation}
\left\{
\begin{aligned}
&H_0(z)=-\sum_{i=1}^N\left[Js_i^zs_{i+1}^z+h_xs_i^x+h_zs_i^z\right], \\
&V(t;z)=-\xi\left[\cos(\omega t)\sum_{i=1}^Ns_i^zs_{i+1}^z+\sin(\omega t)\sum_{i=1}^Ns_i^x\right],
\end{aligned}
\right.
\end{equation}
the high-frequency expansion of the Floquet Hamiltonian is given by $H_F=H_0+\Omega_1/\omega+\Omega_2/\omega^2+\dots$ with
\begin{align}
&\frac{1}{\omega}\Omega_1=-\frac{\xi^2}{\omega}\sum_{i=1}^N\left(s_i^ys_{i+1}^z+s_i^zs_{i+1}^y\right), \\
&\frac{1}{\omega^2}\Omega_2=\frac{\xi^2}{\omega^2}\sum_{i=1}^N\left[h_x\left(s_i^x(s_{i+1}^z)^2+(s_i^z)^2s_{i+1}^x+2s_{i-1}^zs_i^xs_{i+1}^z\right)+2J(s_i^zs_{i+1}^z-s_i^ys_{i+1}^y)+h_zs_i^z\right].
\end{align}

The driving field in the dressed Hamiltonian is expressed as $V^{(n)}(t)=\sum_mV^{(n)}_me^{-im\omega t}$.
Since only the terms of $m=\pm 1$ give dominant contributions to the heating rate, we approximate $V^{(n)}(t)\approx V^{(n)}_1e^{-i\omega t}+V^{(n)}_{-1}e^{i\omega t}$ with $V^{(n)}_{-1}=V^{(n)*}_1$.
It is given by
\begin{align}
&V^{(1)}(t)\approx\frac{i\xi}{\omega}\sum_{i=1}^N\left[(h_x+iJ)(s_i^ys_{i+1}^z+s_i^zs_{i+1}^y)+ih_zs_i^y\right]e^{i\omega t}+\text{(c.c)}, \\
&V^{(2)}(t)\approx\frac{2\xi}{\omega^2}\sum_{i=1}^N\left[(h_x+iJ)J\left(s_i^x(s_{i+1}^z)^2+(s_i^z)^2s_{i+1}^x+2s_{i-1}^zs_i^xs_{i+1}^z\right)+2(h_x+iJ)h_x(s_i^ys_{i+1}^y-s_i^zs_{i+1}^z)\right.\nonumber \\
&\qquad\qquad+(h_x+2iJ)h_z(s_i^xs_{i+1}^z+s_i^zs_{i+1}^x)-ih_xh_zs_i^z+ih_z^2s_i^x\Big]e^{i\omega t} \nonumber \\
&\qquad\qquad-\frac{7i\xi^3}{12\omega^2}\sum_{i=1}^N\left[s_i^x(s_{i+1}^z)^2+(s_i^z)^2s_{i+1}^x+2s_{i-1}^zs_i^xs_{i+1}^z-2i(s_i^ys_{i+1}^y-s_i^zs_{i+1}^z)\right]e^{i\omega t} + \text{(c.c.)}.
\end{align}

Next, we consider the quantum model
\begin{equation}
\left\{
\begin{aligned}
&H_0=-\sum_{i=1}^N[J_z\sigma_i^z\sigma_{i+1}^z+J_x\sigma_i^x\sigma_{i+1}^x+h\sigma_i^z], \\
&V(t)=-\xi\sgn[\cos(\omega t)]\sum_{i=1}^N\sigma_i^x.
\end{aligned}
\right.
\end{equation}
The high-frequency expansion of $H_F=H_0+\Omega_1/\omega+\Omega_2/\omega^2+\dots$ is given by
\begin{align}
&\frac{1}{\omega}\Omega_1=0, \\
&\frac{1}{\omega^2}\Omega_2=\frac{\pi^2\xi^2}{6\omega^2}\sum_{i=1}^N\left[2J_z(\sigma_i^z\sigma_{i+1}^z-\sigma_i^y\sigma_{i+1}^y)+h\sigma_i^z\right].
\end{align}
The driving field in the dressed Hamiltonian, $V^{(n)}(t)\approx V^{(n)}_1e^{-i\omega t}+V^{(n)}_{-1}e^{i\omega t}$ is given by
\begin{align}
&V^{(1)}(t)=\sum_{m\neq 0}\frac{[H_0,V_m]}{m\hbar\omega}e^{-im\omega t}\label{eq:V1} \\
&V^{(2)}(t)\approx\frac{8i\xi}{\pi\omega^2}\sum_{i=1}^N\left[2J_z^2\sigma_{i-1}^z\sigma_i^x\sigma_{i+1}^z
+J_zJ_x(\sigma_{i-1}^y\sigma_i^y\sigma_{i+1}^x+\sigma_{i-1}^x\sigma_i^y\sigma_{i+1}^y-\sigma_{i-1}^z\sigma_i^z\sigma_{i+1}^x-\sigma_{i-1}^x\sigma_i^z\sigma_{i+1}^z)
\right.\nonumber \\
&\qquad\qquad\left.+h_z(2J_z-J_x)(\sigma_i^x\sigma_{i+1}^z+\sigma_i^z\sigma_{i+1}^x)+(2J_z^2+h_z^2)\sigma_i^x\right]e^{i\omega t}+\text{(c.c)}.
\end{align}

It should be noted that the heating rate for $n=1$ evaluated by using the formula~(\ref{eq:quantum_kappa}) is exactly identical to that for $n=0$.
It is shown as follows.
The first-order dressed Hamiltonian is given by $\tilde{H}^{(1)}(t)=H_0+V^{(1)}(t)$ with Eq.~(\ref{eq:V1}).
In Eq.~(\ref{eq:quantum_kappa}), matrix elements $\braket{b|[H_0,V_m]|a}$ appear, which gives $m\hbar\omega\braket{b|V_m|a}$ because of the condition $E_b-E_a=m\hbar\omega$.
The factor $m\hbar\omega$ cancels by the denominator of Eq.~(\ref{eq:V1}), and hence $[H_0,V_m]/m\hbar\omega$ in Eq.~(\ref{eq:V1}) can be replaced by $V_m$.
Consequently, as for the calculation of the heating rate, the first-order dressed Hamiltonian $\tilde{H}^{(1)}(t)=H_0+V^{(1)}(t)$ is equivalent to the bare Hamiltonian $H(t)=H_0+V(t)$, and hence the heating-rate formula for $n=1$ is exactly identical to that for $n=0$ in the present model.

\section{Details on numerical calculations}

Our formula on the heating rate presented in the main text corresponds to the linear response formula [Eq.~(\ref{eq:quantum_formula}) for a quantum system and Eq.~(\ref{eq:classical_formula}) for a classical system] with the replacement of $H_0\to H_F^{(n)}$ and $V(t)\to V^{(n)}(t)$. 

\subsection*{Quantum systems}
For numerical calculations, we used Eq.~(\ref{eq:quantum_kappa}) with the microcanonical distribution
\begin{equation}
p_a=\left\{
\begin{aligned}
&\frac{1}{\Sigma(E)}& &\text{when }E_a\in[E-\Delta E,E], \\
&0& &\text{otherwise},
\end{aligned}
\right.
\end{equation}
where $E_a$ and $\ket{a}$ are an eigenvalue and the corresponding eigenstate of $H_F^{(n)}$, and $\Sigma(E)$ is the number of eigenstates of $H_F^{(n)}$ with an eigenvalue between $E-\Delta E$ and $E$.

We fix the inverse temperature $\beta$ ($\beta=0.23$ in our numerical calculation presented in the main text) and calculate $E=\sum_aE_ae^{-\beta E_a}/Z$, where $Z=\sum_ae^{-\beta E_a}$.
This energy $E$ is used for the microcanonical distribution.
The energy width $\Delta E$ is fixed as $\Delta E=0.1N$ in numerical calculations.

In Eq.~(\ref{eq:quantum_kappa}), the delta function appears.
In numerics, the delta function is approximated as
\begin{equation}
\delta(x)\approx\left\{\begin{aligned}
&\frac{1}{\delta E}& &\text{if }|x|\leq\frac{\delta E}{2}, \\
&0& &\text{otherwise}.
\end{aligned}
\right.
\end{equation}

\subsection*{Classical systems}
Since the Fourier transform of the auto-correlation function of $\dot{V}^{(n)}(z_t)$ equals that of $V^{(n)}(z_t)$ multiplied by $\omega^2$, we used the following formula to calculate the heating rate:
\begin{equation}
\kappa=\frac{\beta}{2N}\sum_{m\neq 0}(m\omega)^2\int_{-\infty}^\infty dt\,\braket{V_m^{(n)*}(z_t)V^{(n)}_m(z_0)}e^{im\omega t}.
\end{equation}
By using the Wiener-Khinchin theorem, the auto-correlation function in the above equation is evaluated through the power spectrum of $V^{(n)}(z_t)$.
The inverse temperature $\beta$ is fixed to be 0.9.

\subsection*{Frequency dependences}

\begin{figure}[h]
\centering
\begin{tabular}{cc}
(a) & (b) \\
\includegraphics[width=0.5\linewidth]{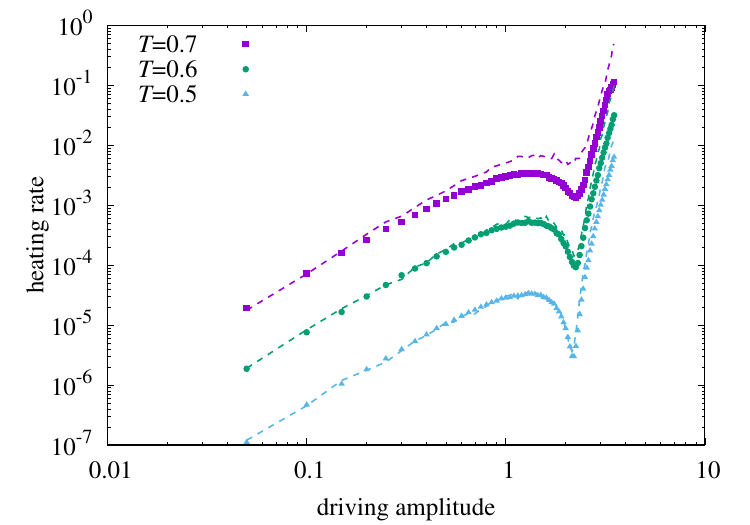}&
\includegraphics[width=0.5\linewidth]{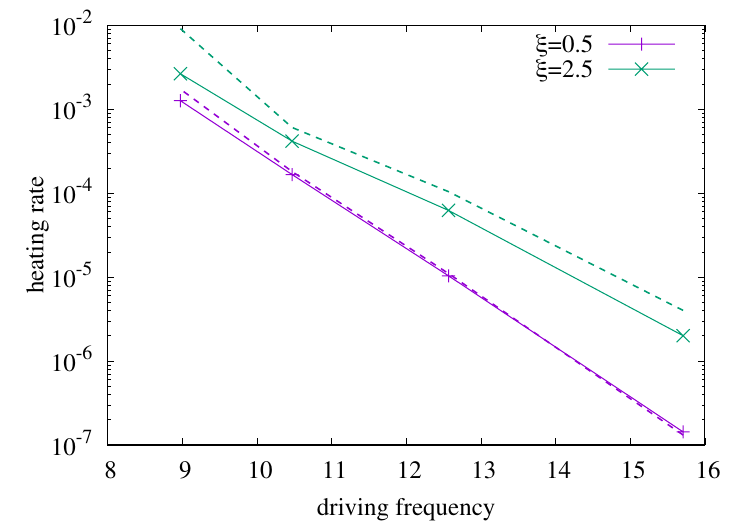}
\end{tabular}
\caption{(a) Heating rates calculated by exactly solving the equations of motion (symbols) and those estimated by our formula with $n=2$ (dashed lines).
Different colors imply different period (frequency) of the driving.
(b) Frequency dependence of the heating rate for weak ($\xi=0.5$) and strong ($\xi=2.5$) driving.
Dashed lines correspond to the heating rate estimated by our formula with $n=2$.}
\label{fig:freq}
\end{figure}

We show numerical results on the heating rate for different frequencies.
In Fig.~\ref{fig:freq} (a), for the same classical model discussed in the main text, we compare the heating rate calculated by solving the equations of motion and that evaluated by our formula with $n=2$.
We see that our formula reproduces the actual heating rate.
For lower frequencies, we see larger deviations, which is consistent with the fact that our formula is based on the high-frequency expansion.

In Fig.~\ref{fig:freq} (b), we show the frequency dependence of the heating rate.
We see that heating is exponentially suppressed at high frequencies.
It is noted that exponential suppression of the heating is rigorously established in quantum lattice systems~\cite{Kuwahara2016, Mori2016b, Abanin2017}.
In Ref.~\cite{Hodson2021}, it is argued that it also occurs in classical systems.
The present numerical result is consistent with those previous studies.

The heating rate at high frequencies thus behaves as $\kappa\sim e^{-c\omega}$ with $c>0$.
From Fig.~\ref{fig:freq}, we see that the coefficient $c$ depends on the driving amplitude $\xi$.

\bibliography{apsrevcontrol,physics.bib}